\documentclass[runningheads]{llncs}
\usepackage{comment}
\usepackage{url}
\usepackage{algorithm,algpseudocode}
\usepackage{amsmath}
\usepackage{amsfonts}
\usepackage{amssymb}
\usepackage{graphicx}
\usepackage{soul}
\usepackage{xcolor}
\usepackage{hyperref}
\usepackage{booktabs}
\usepackage{multirow}
\usepackage{siunitx}

\authorrunning{Abaid Z., Shaghaghi A., et al.}
\titlerunning{Health Access Broker (HAB)}

\newcommand\Tstrut{\rule{0pt}{2.6ex}} 

\begin{document}

\title{Health Access Broker: Secure, Patient-Controlled Management of Personal Health Records in the Cloud}

\author{Zainab Abaid\inst{1} \and
Arash Shaghaghi\inst{2,1} \and
Ravin Gunawardena\inst{4,5} \and
Suranga Seneviratne\inst{4} \and Aruna Seneviratne\inst{1, 3} \and Sanjay Jha\inst{1}}
\institute{The University of New South Wales (UNSW), Sydney, Australia \and
Centre for Cyber Security Research and Innovation, Deakin University, Australia \and Data61, CSIRO, Australia \and The University of Sydney, Australia \and
University of Moratuwa, Sri Lanka \\
\email{z.abaid@unsw.edu.au, a.shaghaghi@deakin.edu.au,ravinsg.16@cse.mrt.ac.lk,
suranga.seneviratne@sydney.edu.au,
\{a.seneviratne, sanjay\}@unsw.edu.au}}

\maketitle

\vspace{-1em}

\begin{abstract}
Secure and privacy-preserving management of Personal Health Records (PHRs) has proved to be a major challenge in modern healthcare. Current solutions generally do not offer patients a choice in where the data is actually stored, and also rely on at least one fully trusted element that patients must also trust with their data. In this work, we present the Health Access Broker (HAB), a patient-controlled service for secure PHR sharing that (a) does not impose a specific storage location (uniquely for a PHR system), and (b) does not assume any of its components to be fully secure against adversarial threats. Instead, HAB introduces a novel auditing and intrusion-detection mechanism where its workflow is securely logged and continuously inspected to provide auditability of data access and quickly detect any intrusions.
\end{abstract}


\section{Introduction}

Modern healthcare requires access to patient data from diverse sources such as smart Internet of Things (IoT) health devices, hospitals, and laboratories. Thus, recent years have seen a move from traditional storage of patient health data  at individual health institutions towards the Personal Health Record (PHR): a more integrated and patient-centered model with a patient's data from different sources all available in one place. Maintaining security and privacy of patient PHRs is an ongoing research problem, given the large number of data providers (e.g. pathologists, paramedics) and consumers (e.g. pharmacists, doctors) who need to access the PHRs in various scenarios.
Currently deployed PHR systems require patients to trust particular commercial providers (e.g. Indivo\footnote{http://indivohealth.org}) or institutions such as the government (e.g. My Health Record\footnote{https://www.myhealthrecord.gov.au} in Australia). Patients have increasingly opted out of these services because of the need to trust in a  particular institution or provider's security and privacy mechanisms, which are complicated and sometimes unknown. Breaches of healthcare data~\cite{liu2015data} have further eroded patients' trust in these systems. 

A common solution presented in prior research is to develop strong cryptographic access control methods for PHRs stored in public clouds~\cite{li2012scalable}. Records can be encrypted by patients before uploading to untrusted cloud services. However, these methods demand significant management and computing overhead from data owners. Moreover, some elements, such as key management servers, are still fully trusted, and users do not have a choice in terms of where to store their data. Thus, the users may not be  comfortable with this solution.

In this paper, we present Health Access Broker (HAB), a secure PHR management system that provides the users with full control of where their data is stored. HAB assumes that any of its components may be compromised, and introduces a novel auditing and intrusion-detection mechanism to handle it. Patients encrypt their PHRs under a secure attribute-based encryption scheme, specify an access policy, and pass them to HAB, which acts as an intermediate service to upload the data to the public cloud services of the patients' choice. The data is split across multiple cloud services as an added layer of security, such that any one share of the data is meaningless on its own. When another user wants to access a patient's record, HAB checks the access policy and retrieves and aggregates the data for the user, who then decrypts it using their key. Access to data is immediate as there is no patient involvement in the retrieval process.

Compared to existing services, HAB is more likely to be trusted by patients for two reasons. Firstly, it does not locally store patient data or users' encryption keys, and therefore cannot decrypt and view the data. Secondly, HAB's workflow and all user-HAB communication is securely logged (e.g. on a private blockchain). To protect against adversarial threats, HAB actions are  continuously compared with client requests to ensure that an adversary (e.g. a system administrator) cannot initiate any data management operations not requested
by an authorised user. Because of continuous inspection of the logs, any other events indicating HAB compromise can be also be detected and handled quickly.

The rest of this paper is organised as follows. Section~\ref{sec:related} outlines related work in the area.  Section~\ref{sec:architecture} describes the system architecture and algorithms, and security model. In Section~\ref{sec:evaluation}, we evaluate performance of a prototype implementation, and Section~\ref{sec:conclusion} concludes the paper. 

\vspace{-0.25em}
\section{Related Work}
\label{sec:related}
\vspace{-0.25em}
Prior research on securing health data in public clouds has taken two kinds of approaches that inspire our current work: (i) access to cloud data via a trusted middleware, and (ii) cryptographically-enforced access control.
 
The first category of work uses a model that has been popular in commercial cloud-computing solutions: the cloud access broker~\cite{nair2010towards}, a trusted middleware between the users and the cloud which manages user interaction with cloud services  and enforces access control. 
For example,
Wu et al.~\cite{wu2012secure} propose the use of a broker service for aggregation of patient health records stored on different clouds by various health providers. 
The focus of their work is an algorithm for aggregating records from different organisations given that each follows a different schema. This makes it orthogonal to our work, in which the key job of the broker service is access control and security of health data.
~\cite{matos2018securing} also implements middleware for managing retrieval and access control for electronic health records (EHRs) on a commercial cloud service. They use existing third-party services
to manage data storage and access control and encryption, and implement the middleware themselves. Thus, users of the system will need to trust these third-party services. In our work, we propose a model 
that does not require trusting a specific provider, as users may store their records anywhere.


Access control based on Attribute-based Encryption (ABE) is a commonly used method in prior work~(e.g., \cite{jahan2017secure,jahan2017light}) due to the fine granular control it offers and its flexibility over traditional access control methods such as Role-Based Access Control (RBAC). For instance, Yu et al.~\cite{yu2010achieving} proposed a system where a key-policy ABE (KP-ABE) scheme is used to encrypt data with a set of attributes before uploading it to the untrusted cloud. Each user possesses a key representing their access policy, and only the users whose access policy matches a file's attributes can decrypt the file. User revocation is handled by the cloud service using proxy re-encryption.

ABE has also been proposed for the eHealth domain~\cite{greene2018secure,liu2015secure,li2012scalable,eom2016patient,qian2015privacy,mubarakali2020design,doshi2019enhanced}.
 Greene et al.~\cite{greene2018secure} use KP-ABE to restrict access to health data, and additionally incorporate hash-chaining for time-based access control; however the scenario is  different from ours and deals with data sharing from smart health devices to a cloud database. 
Narayan et al.~\cite{narayan2010privacy} use ciphertext-policy ABE (CP-ABE) to secure PHRs in a centralised cloud server and also allow for keyword search over encrypted records. ~\cite{xhafa2015designing} improves Narayan et al.'s work by adding fault tolerance, efficient local decryption, and making the keyword search much more lightweight. A key drawback of both schemes is the heavy workload expected from the patient, who needs to take care of access grants/revokes, data uploads, and update approvals. In our work, we propose a broker service that manages most of these tasks; the minimal set of tasks a patient needs to perform can be done through an easy-to-use mobile application.


\section{Health Access Broker}
\label{sec:architecture}

\begin{figure*}[t]
    \centering
    \includegraphics[width=0.8\textwidth]{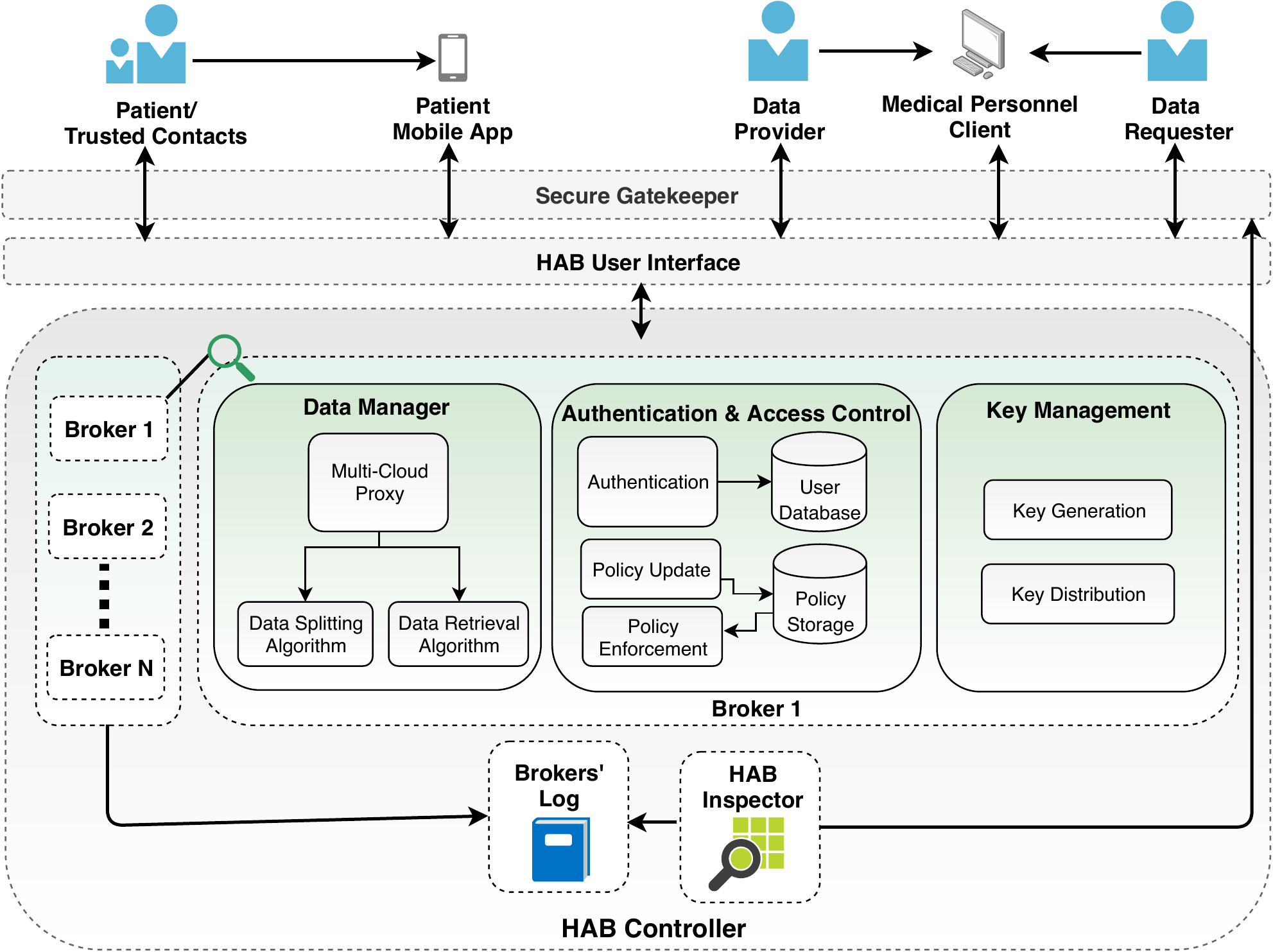}
    \caption{Health Access Broker architecture}
    \label{fig:arch}
\end{figure*}


Our proposed architecture is shown in Figure~\ref{fig:arch}. There are three categories of HAB users: 
\textit{patients and trusted contacts} (e.g. primary carers -- henceforth referred to as just patients), who control the data that is entered into a patient's health record and determine access policies for it; \textit{Data Requestors} (DRs), which may include doctors, analysts, or anyone wishing to view a patient's data; and \textit{Data Providers} (DPs), usually medical personnel that may add to a patient's health record.
Users interact with the HAB controller through a web-based user interface (UI). In addition, there is a mobile application for patients and a desktop application for medical personnel that must be used to perform particular tasks which cannot be directly performed on the web-based UI for security reasons. 
To avoid a single point of failure and for quick fault localisation, HAB comprises multiple brokers, each of which contains the same functionality but will in practice be deployed at a different server responsible for a different set of patients. Each user connecting to the HAB controller through the UI will be allocated to the relevant broker which will handle all data operations requested by the user.

Each broker comprises the following three modules: 
\begin{enumerate}
    \item The Data Storage, Retrieval And Update Module is responsible for
    indexing and storing data with external cloud providers as well as retrieving portions of data requested by users.
    \item The Authentication and Access Control Module is responsible for managing user authentication and access policies, including storing policy specifications, enforcing policies, updating policies and user access revocation.
    \item The Key Management Module manages key generation, distribution and update, and stores the keys used for the encryption scheme under which patient data is protected. 
\end{enumerate}

A HAB Inspector (HI) that is external to the brokers ensures HAB security based on two components, a  
\textit{Gatekeeper} and a \textit{Brokers' Log} (details below).


\subsection{The HAB Workflow}\label{sec:habprotocol}
HAB's workflow can be divided into four key functions that we describe below.

\paragraph{User Registration and Authentication}
A first-time HAB user initiates a registration request through the HAB UI and chooses a username and password combination for future authentication. The UI assigns the user to a particular broker. The broker invokes its Key Management Module (KMM) to issue a key to the user. Existing users enter their username and password into the UI, which assigns them to a particular broker. The broker invokes its Authentication and Access Control Module (AACM) to authenticate the user.

\paragraph{Data Upload and Update}
A DP who wants to upload data enters the patient's identifying information and the data itself into a desktop application, which encrypts the data and sends it to HAB. HAB sends the data to the patient's mobile app for review. The patient reviews and approves the data, sets an access policy for it, and specifies a list of cloud-based storage locations for it. The app encrypts it under an encryption scheme using the access policy and sends the encrypted data and its access policy back to HAB. The data is received by the Data Manager Module (DMM) of the broker that the patient is assigned to. The DMM passes the data to the Multi-Cloud Proxy (MCP), which in turn runs a data splitting algorithm and stores each chunk of data in one of the storage locations chosen by the patient. The access policy is passed to the Access Control and Authentication Module (AACM) for storage. Data update process is very similar, except that it begins with an additional searching step to retrieve the portion of a health record that is to be updated. 
Although data upload/update will be delayed until patient approval, this is to ensure data integrity. The delay is acceptable because data upload is not usually an urgent matter.

\paragraph{Data Retrieval}
An authenticated Data Requestor (DR) makes a request for a particular data item belonging to a particular patient, which is passed to the DMM. The DMM invokes the AACM, which checks whether the DR satisfies the access policy specified by the patient for the requested data. If the check is successful, the DMM requests the MCP for the required data, which in turn runs the retrieval algorithm to combine the chunks of data stored across multiple clouds and returns the data to the DMM.
If the check is unsuccessful, the DR may send a request to the patient to update their access policy and allow access. 

HAB would also provide emergency access for hospital emergency rooms (ERs) urgently requiring a patient's data. We propose using a symmetric encryption scheme where all hospitals using HAB possess an emergency public/private key pair. When a hospital ER requests a patient's data, HAB uses proxy re-encryption to re-encrypt it with the hospital's public key. The ER staff can then decrypt the received data with the hospital's private key. 
\vspace{-0.25cm}
\paragraph{Auditing and Intrusion Detection}
Existing work in securing cloud-based PHRs generally assumes at least one fully trusted element, for key management or other important functions. Our work is set apart by the fact that while HAB is a trusted entity, it is strictly audited to protect against compromise. HAB's auditing mechanism comprises a \textit{Gatekeeper}, a \textit{Brokers' Log} (BL) and a \textit{HAB Inspector} (HI). The
Gatekeeper monitors all communication between a connected user and a broker, logging all requests made by a user along with the user's identifying information and timestamps. The BL is a secure log of all the actions performed by HAB modules in response to user requests, e.g. data storage or policy update. The HI runs continuously and matches the Gatekeeper's logs with the BL, using a set of intrusion detection rules which specify the fields of the logs that are to be compared for each event. For example, for a data retrieve event, it matches the file ID being requested with the file ID being retrieved by HAB. This is to identify any actions performed by a compromised broker that were not requested by an authorised user, for example, retrieving additional data to what is requested. The patient and system admin are both alerted if an instrusion is detected.
We propose that the BL is maintained in a private blockchain accessed by all brokers in HAB, to ensure its immutability.

\subsection{HAB Algorithms}
We now outline the specific algorithms or schemes we propose for the following HAB functions: data splitting for upload, data aggregation for retrieval, and encryption and revocation. In a practical deployment, HAB's modular design allows for easy modification of the algorithms for any of these functions without affecting the rest of the functionality.

\vspace{-0.5em}
\paragraph{Data Splitting}
We implement a highly secure data storage algorithm where a document is first encrypted and then split into $N$ chunks. $N$ is a configurable parameter representing the number of cloud services that will store the document. Each chunk is stored in a different cloud service. We split the document according to Shamir's secret-sharing scheme as an added layer of security. A secret shared under this scheme cannot be reconstructed unless a configurable threshold $T$ of the $N$ shares become known; any one of the shares is meaningless on its own. Thus, at least $T$ of the $N$ cloud services 
will need to collude (or be compromised by an adversary)
for reconstruction of the encrypted version of a document. This scheme therefore represents an added layer of security
in case of broken encryption or key compromise.
Algorithm~\ref{alg:split} describes how we apply Shamir's scheme to a file and then store chunks of the file in $N$ clouds.

\begin{algorithm}[!t] 
  \footnotesize
\caption{Splitting and Uploading File to Clouds}
\label{alg:split}
\begin{algorithmic}[1]
\Require{Number of clouds $N$, Threshold $T$, Data File $F$} 
\Ensure{Chunks $N_i$ stored in clouds $C_i$}
\State {Run secret sharing scheme with threshold $T$ on file $F$ to generate shares $S_i$ where $i = 1,2, \dots N$}
    \For{Share $S_i$}                    \State {Label $S_i$ with File ID and Share ID}
        \State {Upload share $S_i$ to cloud $C_i$}
        \State {Store [File ID, Share ID, $C_i$] in index table}
    \EndFor
\end{algorithmic}
\end{algorithm}

\vspace{-0.5em}
\paragraph{Data Aggregation for Retrieval}
Once the DMM obtains authorisation from the AACM to retrieve a file, it runs Algorithm~\ref{alg:retrieve} for the retrieval, which retrieves the chunks of the file from the $N$ cloud services, and then runs the \textit{combine} algorithm of Shamir's secret sharing scheme to reconstruct the original file. 
\vspace{-0.5cm}
\begin{algorithm} 
 \footnotesize
\caption{Reconstructing a File Retrieved from Clouds $C_i$ where $i = 1,2, \dots ,N$ }
\label{alg:retrieve}
\begin{algorithmic}[1]
\Require{Threshold $T$, Shares $S_i$ of File $F$ where $i = 1,2, \dots ,T$, File index table} 
\Ensure{File F reconstructed from Shares $1,2 \dots ,N$ stored in Clouds $1,2 \dots ,N$}
\State {From File index table, retrieve Pointers $P_i$ to cloud locations of $T$ shares of File $F$, where $i = 1,2, \dots ,T$}
    \For{Each Pointer $P_i$}                    \State {Retrieve share $S_i$ from Cloud $C_i$  according to Pointer $P_i$}
        \State {Reconstructed file $R$ = $R$ + $S_i$}
    \EndFor
    \State {Run secret-combining algorithm on $R$ to obtain File F}
\end{algorithmic}
\end{algorithm}

\vspace{-0.5em}
\paragraph{Encryption}

In PHR applications, the access policy for the data is usually known at the time of creation of the data, while the identities of the recipients are not necessarily known. We therefore propose using a Ciphertext-policy Attribute-based Encryption (CP-ABE) scheme where the attributes of users (e.g.  organisation and department) serve as their keys, and the data owner can define an access policy in terms of the minimum set of attributes a user must possess in order to be able to decrypt a data file. We currently use the CP-ABE scheme presented in~\cite{bethencourt2007ciphertext}; a formal definition is omitted owing to space constraints.

A key requirement in PHR encryption is efficient user revocation, i.e. denial of access to previously authorised users. Most revocation schemes for CP-ABE rely either on proxy re-encryption, which has a large computational overhead, or time-based keys, in which case the revocation is not immediate~\cite{liu2014time}. Instead, inspired by \textit{mediated} CP-ABE~\cite{ibraimi2009mediated}, which allows for immediate revocation by involving a semi-trusted server in the decryption process, we use the following simple revocation mechanism. Patients can associate a list of revoked users with each data item in their health record, as well as a list of users who cannot view any of their data. When a user makes a data retrieval request for a patient, HAB (i.e. the AACM) first checks that the user's access is not revoked generally or for the requested item, and only then begins the retrieval process. Thus, as our system already involves an intermediary in data retrieval, we are able to augment any encryption scheme with immediate and efficient revocation without requiring any re-encryption or key updates.

 For attribute assignment and key generation, we assume that the healthcare system can be divided into domains when HAB is practically deployed. Each domain functions as an \textit{attribute issuing authority} (AIA) that
is responsible for issuing attribute sets(e.g. hospital name, department, specialisation) to the users in its domain. Each user applies to its AIA for assignment of attributes. Attributes are drawn from an (extendable) attribute universe that can be defined as part of a HAB deployment. HAB's Key Management Module receives these attributes directly from the AIA through a secure channel when a new user applies to register with HAB, and issues a key based on them.
\vspace{-1em}
\subsection{HAB Security}
\label{sec:model}

\setlength{\tabcolsep}{0.75em}
\begin{table}[t]
\fontsize{8.0}{9.0}\selectfont 
\begin{tabular}{p{5.75cm}p{5.75cm}}
\toprule
Adversarial Threat & Design Decision\\ 
\midrule 
(1-A) Adversary performs password attacks to obtain credentials of user authorised to view a patient's data, and gains access to data that he is otherwise not authorised to view. & A practical deployment will enforce strong passwords and limit the number of login attempts. In future, we can add anomaly detection to HAB Inspector and compare a user's actions with their normal profile to detect account misuse. \\
\Tstrut (1-B) Adversary obtains secret key of an authorised user and uses it to decrypt data.  
& HAB does not store issued keys; standard network defenses~\cite{conti2016survey} would be applied in practice to defend against man-in-the-middle attacks during key distribution. 
\\
\Tstrut (1-C) Two adversaries combine their keys to decrypt data that neither can decrypt individually. & HAB's CP-ABE scheme is collusion-resistant~\cite{bethencourt2007ciphertext}.\\
\Tstrut (1-D) Adversary compromises, or colludes with, the cloud service storing a patient's data to directly obtain patient data bypassing HAB's access control. & Each data item is split across multiple clouds and cannot be reconstructed until a (configurable) number of shares of the item are obtained; the adversary cannot realistically compromise several cloud services.\\  
\Tstrut (2-A) Malicious data provider (DP) deliberately sends inaccurate/false data for upload into patient's health record. & No data can be entered into a patient's health record without the patient's review and approval.\\
\Tstrut (2-B) An authorised but malicious data provider (DP) deliberately enters false information in an existing record. & Updates must also be patient-approved.\\
 
\Tstrut (3-A) Adversary performs Denial-of-Service (DoS) attacks on HAB controller. & In a practical deployment, existing well-known defenses against DoS attacks (e.g. \cite{jazi2017detecting}) should be used to secure HAB servers.\\
\Tstrut (3-B) Adversary obtains a patient's credentials (e.g. by password attacks or social engineering) and modifies access policy of files to deny access to previously authorised users. & Same as (1-A)\\ \bottomrule
\end{tabular}
\vspace{0.5em}
\caption{Adversarial threats and corresponding defenses.}
\label{tab:adversary_model}
\vspace{-2em}
\end{table}

HAB is designed to protect the confidentiality, integrity and availability of patient health data based on a \textit{secure-by-design} philosophy. Thus, we take adversarial threats into account while defining its basic protocols.  Table~\ref{tab:adversary_model} summarises these threats. Threats (1-A) to (1-D) correspond to data confidentiality issues; (2-A) and (2-B) correspond to integrity, and (3-A) and (3-B) correspond to availability. The table also outlines design decisions to defend against these threats; note that some defenses are to be dealt with in the deployment rather than the design stage and are left to a practical deployment. 

Some security threats may arise because of users' lack of security-awareness. Passwords may be guessed by adversaries or obtained by social engineering attacks, or users may inadvertently set the wrong access policies and allow access to unintended persons. The first threat can be minimised by requiring strong passwords and through user education. The second can be handled by designing the patient application such that when setting the access policy for a data item, no option of a “public” or “allow-all” access policy is provided which can be inadvertently selected by inexperienced users; rather, patients need to specify particular user attributes required for gaining access.

\section{Prototype Implementation and Evaluation}
\vspace{-0.5cm}
\label{sec:evaluation}
\setlength{\tabcolsep}{0.75em}
\begin{table}
\label{tab:results}
  \centering
  \begin{tabular}{lSSSSSS}
    \toprule
    \multirow{2}{*}{File Size} &
      \multicolumn{2}{c}{Splitting (s)} &
      \multicolumn{2}{c}{Encryption (s)} &
      \multicolumn{2}{c}{Upload (s)} \\
      & {Average} & {Std. Dev} & {Average} & {Std. Dev} & {Average} & {Std. Dev} \\
      \midrule
    1 KB & 1.5 & 0.3 & 1.2 & 0.5& 19.3 & 2.2 \\
    10 KB & 3.0 & 0.7 & 1.0 & 0.3 & 22.4 & 2.8 \\
    100 KB & 27.7 & 6.8 & 1.0 & 0.3 & 69.9 & 2.3 \\
    500 KB & 124.0 & 15.3 & 0.8 & 0.1 & 212.1 & 6.8 \\
    1 MB & 363.0 & 23.7 & 1.1 & 0.3 & 490.0 & 7.9 \\
    \bottomrule
  \end{tabular}
  \vspace{0.25em}
\caption{Running time (in seconds) of HAB operations for different file sizes.}
  \vspace{-2em}
\end{table}

We have implemented a small-scale prototype of HAB and evaluate the latency of common operations.
We used Django\footnote{https://www.djangoproject.com/} to implement the HAB UI as a web application and deployed it to a remote\footnote{www.pythonanywhere.com} webserver. 
We set up HTML forms in the frontend of the application to provide the basic functionality of data upload and retrieval (we used Google Drive in our prototype evaluation), and integrated access control and policy management functions into the Django backend that integrates an SQLite database. We also wrote Java-based clients to interact with it to simulate the patient's mobile app and medical personnel's client app; the encryption/decryption and data splitting functionality were part of these clients. By running the clients after deploying the server, we tested the latency of common operations for different sizes of data items. We used random XML files as data items, as HAB in deployment will use a standardised XML-based health data exchange format such as HL7 CDA~\cite{dolin2006hl7}. 

The results for the operations whose running times vary by file size are summarised in Table 2. The data splitting and encryption operations take place offline (i.e. on the client device), while the file upload process involves server interaction.
The upload process took approximately 25 seconds for smaller files ($<$100 KB) and 43.8 seconds for a 1 MB file; the relatively large running time is due to authentication and file upload to external cloud services, as well as several database operations to properly index the uploaded file on the HAB backend. However, as data upload is generally not urgent, this upload time is acceptable.  For large files, the data splitting process is the most time-consuming operation, as its running time grows significantly with file size; a 1 KB file required 0.41 seconds to split, but a 1 MB file required 384 seconds. However, this is not a significant problem for the following reasons. Firstly, as mentioned above, data upload is generally not urgent. Secondly, the splitting operation can be made optional if file sizes are always large, as the splitting process is just an added layer of security; if the patient wishes to speed up data upload, they can opt out of data splitting and choose instead to store the entire file with a single cloud service. Thirdly, in practical settings, apart from images such as ultrasound or X-Ray films, most files uploaded will be short text-based documents of a few pages and thus minimal ($<$100 KB) in size. 

We also tested two revocation operations: first, when a patient wishes to deny access to their files for a particular user (user revocation), and second, for a set of users owning a particular attribute (attribute revocation). Both these operations were completed in only 0.01 seconds.
Finally, we tested the running time of the policy update operation, i.e. if a patient sends a new access policy to be stored against a particular file. This was completed in 0.74 seconds. 


\section{Conclusion}
\label{sec:conclusion}
In this paper, we presented the Health Access Broker (HAB), a PHR-sharing platform that allows patients total control over their data in terms of where it is stored and who can access it. Compared to existing solutions, HAB does not assume any component to be fully trusted or resilient to compromise. Instead, it deals with possible adversarial threats with a novel auditing and intrusion-detection mechanism based on continuous logging and inspection of logs. Based on a prototype implementation, we have shown that common workflows can be completed in time-frames suitable for real-time and user friendly operation. In future, HAB's capabilities can be further extended by incorporating anomaly detection to detect adversarial use of legitimate user profiles. We will be also exploring leveraging granular access control solutions such as Function-based Access Control (FBAC)\cite{Desmedt2018} for HAB.

\addcontentsline{toc}{section}{Bibliography}
\newpage
\bibliographystyle{abbrv} 
\bibliography{health_bibliography}

\begin{thebibliography}{10}

\bibitem{bethencourt2007ciphertext}
J.~Bethencourt, A.~Sahai, and B.~Waters.
\newblock Ciphertext-policy attribute-based encryption.
\newblock In {\em 2007 IEEE symposium on security and privacy (SP'07)}, pages
  321--334. IEEE, 2007.

\bibitem{conti2016survey}
M.~Conti, N.~Dragoni, and V.~Lesyk.
\newblock A survey of man in the middle attacks.
\newblock {\em IEEE Communications Surveys \& Tutorials}, 18(3):2027--2051,
  2016.

\bibitem{Desmedt2018}
Y.~Desmedt and A.~Shaghaghi.
\newblock {\em Function-Based Access Control (FBAC): Towards Preventing Insider
  Threats in Organizations}, pages 143--165.
\newblock Springer International Publishing, Cham, 2018.

\bibitem{dolin2006hl7}
R.~H. Dolin, L.~Alschuler, S.~Boyer, C.~Beebe, F.~M. Behlen, P.~V. Biron, and
  A.~Shabo.
\newblock Hl7 clinical document architecture, release 2.
\newblock {\em Journal of the American Medical Informatics Association},
  13(1):30--39, 2006.

\bibitem{doshi2019enhanced}
N.~Doshi, M.~Oza, and N.~Gorasia.
\newblock An enhanced scheme for phr on cloud servers using cp-abe.
\newblock In {\em Information and Communication Technology for Competitive
  Strategies}, pages 439--446. Springer, 2019.

\bibitem{eom2016patient}
J.~Eom, D.~H. Lee, and K.~Lee.
\newblock Patient-controlled attribute-based encryption for secure electronic
  health records system.
\newblock {\em Journal of medical systems}, 40(12):253, 2016.

\bibitem{greene2018secure}
E.~Greene, P.~Proctor, and D.~Kotz.
\newblock Secure sharing of mhealth data streams through
  cryptographically-enforced access control.
\newblock {\em Smart Health}, 2018.

\bibitem{ibraimi2009mediated}
L.~Ibraimi, M.~Petkovic, S.~Nikova, P.~Hartel, and W.~Jonker.
\newblock Mediated ciphertext-policy attribute-based encryption and its
  application.
\newblock In {\em International Workshop on Information Security Applications},
  pages 309--323. Springer, 2009.

\bibitem{jahan2017light}
M.~Jahan, M.~Rezvani, Q.~Zhao, P.~S. Roy, K.~Sakurai, A.~Seneviratne, and
  S.~Jha.
\newblock Light weight write mechanism for cloud data.
\newblock {\em IEEE Transactions on Parallel and Distributed Systems},
  29(5):1131--1146, 2017.

\bibitem{jahan2017secure}
M.~Jahan, P.~S. Roy, K.~Sakurai, A.~Seneviratne, and S.~Jha.
\newblock Secure and light weight fine-grained access mechanism for outsourced
  data.
\newblock In {\em 2017 IEEE Trustcom/BigDataSE/ICESS}, pages 201--209. IEEE,
  2017.

\bibitem{jazi2017detecting}
H.~H. Jazi, H.~Gonzalez, N.~Stakhanova, and A.~A. Ghorbani.
\newblock Detecting http-based application layer dos attacks on web servers in
  the presence of sampling.
\newblock {\em Computer Networks}, 121:25--36, 2017.

\bibitem{li2012scalable}
M.~Li, S.~Yu, Y.~Zheng, K.~Ren, and W.~Lou.
\newblock Scalable and secure sharing of personal health records in cloud
  computing using attribute-based encryption.
\newblock {\em IEEE transactions on parallel and distributed systems},
  24(1):131--143, 2012.

\bibitem{liu2015secure}
J.~Liu, X.~Huang, and J.~K. Liu.
\newblock Secure sharing of personal health records in cloud computing:
  Ciphertext-policy attribute-based signcryption.
\newblock {\em Future Generation Computer Systems}, 52:67--76, 2015.

\bibitem{liu2014time}
Q.~Liu, G.~Wang, and J.~Wu.
\newblock Time-based proxy re-encryption scheme for secure data sharing in a
  cloud environment.
\newblock {\em Information sciences}, 258:355--370, 2014.

\bibitem{liu2015data}
V.~Liu, M.~A. Musen, and T.~Chou.
\newblock Data breaches of protected health information in the united states.
\newblock {\em Jama}, 313(14):1471--1473, 2015.

\bibitem{matos2018securing}
D.~R. Matos, M.~L. Pardal, P.~Ad{\~a}o, A.~R. Silva, and M.~Correia.
\newblock Securing electronic health records in the cloud.
\newblock In {\em Proceedings of the 1st Workshop on Privacy by Design in
  Distributed Systems}, page~1. ACM, 2018.

\bibitem{mubarakali2020design}
A.~Mubarakali, M.~Ashwin, D.~Mavaluru, and A.~D. Kumar.
\newblock Design an attribute based health record protection algorithm for
  healthcare services in cloud environment.
\newblock {\em Multimedia Tools and Applications}, 79(5):3943--3956, 2020.

\bibitem{nair2010towards}
S.~K. Nair, S.~Porwal, T.~Dimitrakos, A.~J. Ferrer, J.~Tordsson, T.~Sharif,
  C.~Sheridan, M.~Rajarajan, and A.~U. Khan.
\newblock Towards secure cloud bursting, brokerage and aggregation.
\newblock In {\em 2010 eighth IEEE European conference on web services}, pages
  189--196. IEEE, 2010.

\bibitem{narayan2010privacy}
S.~Narayan, M.~Gagn{\'e}, and R.~Safavi-Naini.
\newblock Privacy preserving ehr system using attribute-based infrastructure.
\newblock In {\em Proceedings of the 2010 ACM workshop on Cloud computing
  security workshop}, pages 47--52. ACM, 2010.

\bibitem{qian2015privacy}
H.~Qian, J.~Li, Y.~Zhang, and J.~Han.
\newblock Privacy-preserving personal health record using multi-authority
  attribute-based encryption with revocation.
\newblock {\em International Journal of Information Security}, 14(6):487--497,
  2015.

\bibitem{wu2012secure}
R.~Wu, G.-J. Ahn, and H.~Hu.
\newblock Secure sharing of electronic health records in clouds.
\newblock In {\em 8th International Conference on Collaborative Computing:
  Networking, Applications and Worksharing (CollaborateCom)}, pages 711--718.
  IEEE, 2012.

\bibitem{xhafa2015designing}
F.~Xhafa, J.~Li, G.~Zhao, J.~Li, X.~Chen, and D.~S. Wong.
\newblock Designing cloud-based electronic health record system with
  attribute-based encryption.
\newblock {\em Multimedia Tools and Applications}, 74(10):3441--3458, 2015.

\bibitem{yu2010achieving}
S.~Yu, C.~Wang, K.~Ren, and W.~Lou.
\newblock Achieving secure, scalable, and fine-grained data access control in
  cloud computing.
\newblock In {\em 2010 Proceedings IEEE INFOCOM}, pages 1--9. Ieee, 2010.

\end{thebibliography}
\end{document}